\documentclass[a4paper,11pt]{article}    % For preprint
\pdfoutput=1 % if your are submitting a pdflatex (i.e. if you have  images in pdf, png or jpg format)

\usepackage{jheppub}
\usepackage{epsfig}
\usepackage[T1]{fontenc} % if needed
\def\arXiv#1{\href{http://arxiv.org/abs/#1}{arXiv:#1}}
\def\arXiv#1#2{\href{http://arxiv.org/abs/#1}{arXiv:#1}}

\def\doi#1#2{\href{http://doi.org/#1}{#2}}

\def\be{\begin{eqnarray}}
\def\ee{\end{eqnarray}}
\def\bea{\begin{eqnarray}}
\def\eea{\end{eqnarray}}
\newcommand{\nn}{\nonumber}
\newcommand\para{\paragraph{}}

\def\lae{\mathrel{\mathop{\smash{\lower .5 ex \hbox{$\stackrel<\sim$}}}}}
\def\lae{\mathrel{\mathop{\smash{\lower .5 ex \hbox{$\stackrel>\sim$}}}}}

\title{\boldmath Momentum relaxation in a holographic Weyl semimetal}

\author{Junkun Zhao}
\affiliation{
Center for Gravitational Physics, Department of Space Science,
\\and International Research Institute
of Multidisciplinary Science,
\\ Beihang University,  Beijing 100191, China}
\emailAdd{junkunzhao@buaa.edu.cn}

\abstract{We study the effects of momentum relaxation on the holographic Weyl semimetal which exhibits a topological quantum phase transition between the Weyl semimetal phase and a topological trivial phase. The conservation of momentum in the field theory is broken by the axion fields in holography. The topological Weyl semimetal phase is characterized by a nontrivial anomalous Hall conductivity. We find that the critical value of the phase transition decreases when we increase the momentum relaxation strength up to a special value, above which it goes to zero. This indicates that the Weyl semimetal phase shrinks and finally disappears as the momentum relaxation strength is increased, which is consistent with the weakly coupled field theory predictions. We also study the behavior of transverse/longitudinal conductivities and low temperature dependence of the d.c.resistivities with respect to momentum relaxation strength.}

\begin{document}
\maketitle
\flushbottom
\pagestyle{plain} \setcounter{page}{1}
\newcounter{bean}
\baselineskip16pt

%%%%%%%%%%%%%%%%%%%%%%%%%%%%%
\section{Introduction}
%%%%%%%%%%%%%%%%%%%%%%%%%%%%%
\para
Weyl semimetal is an interesting and important gapless state of matter in which the low-energy excitations can be described by the Weyl equation. The system inherits lots of exotic transport properties due to the chiral anomaly, which has attracted lots of theoretical and experimental interest \cite{vishwanath,burkov0,Landsteiner:2016led}. As a topological quantum matter, the description of Weyl semimetal goes beyond the Landau-Ginzburg paradigm in terms of symmetry breaking. The same as graphene \cite{jan}, the effective fine structure constant is very large due to the smallness of the Fermi velocity compared to the speed of light. This means that the Weyl semimetal can exist in a strongly interacting region with no quasiparticles, where the perturbed quantum field theory and topological band theory description break down \cite{Gonzalez:2015tsa}. Therefore, it is an important and challenging question to find a proper theoretical description of the strongly coupled Weyl semimetal.

\para
Holographic duality (or AdS/CFT correspondence) relates the $d$-dimensional strongly coupled field theory to a $d+1$-dimensional weakly coupled classical gravitational theory, which is a powerful tool to tackle problems arising in field theory. This method has been applied to solve various problems in condensed matter physics and yielded invaluable insights \cite{Zaanen:2015oix,book0,review}. Recently, the holographic model of strongly coupled Weyl semimetal has been constructed in Refs. \cite{Landsteiner:2015pdh,Landsteiner:2015lsa}, where the Weyl semimetal phase is characterized by a nonzero anomalous Hall conductivity. The system undergoes a topological quantum phase transition from the Weyl semimeatl phase to a topological trivial phase with vanishing anomalous Hall conductivity. Since then, many issues related to holographic Weyl semimetal have been studied, including odd viscosity \cite{Landsteiner:2016stv}, surface state \cite{Ammon:2016mwa}, optical conductivity \cite{Grignani:2016wyz}, axial Hall conductivity \cite{Copetti:2016ewq}, topological invariants \cite{Liu:2018djq}, and nodal line semimetal \cite{Liu:2018bye,Liu:2020ymx}. Other studies can be found in Refs. \cite{Gursoy:2012ie,Hashimoto:2016ize,Ammon:2018wzb,Baggioli:2018afg,Liu:2018spp,Ji:2019pxx,Song:2019asj,
Tanaka:2020yax,Juricic:2020sgg,Baggioli:2020cld,Fadafan:2020fod}, and see \cite{Landsteiner:2019kxb} for a recent review on this topic.

\para
So far, the investigations of the holographic Weyl semimetal are mainly focused on translational invariant systems where the momentum is conserved. In real materials, the momentum of electrons is dissipated due to scattering with the background ion lattice or disorder. In the weakly coupled field theory, the Weyl points can be destroyed by breaking the translational symmetry \cite{Hosur:2013kxa}, which means that the momentum relaxation may have nontrivial physical effects on the properties of the Weyl semimetal. At a strongly coupled region, the Weyl semimetal still exists, and it is important to explore the effects of momentum relaxation on the system \cite{Landsteiner:2015lsa,Landsteiner:2015pdh}. This motivates us to study momentum relaxation in the holographic Weyl semimetal by breaking the translational symmetry along the spatial directions.

\para
We will use the linear axion models \cite{Andrade:2013gsa,Baggioli:2021xuv} to implement the momentum dissipation in the holographic Weyl semimetal. This enables us to break the translational symmetry while retaining the homogeneity of the background geometry. We will focus on low-temperature physics. The reason is twofold. First, the zero-temperature ground state is difficult to construct in the presence of the axion fields. Second, the absolute zero temperature cannot be physically reached in experiments. Because of the existence of the quantum critical region, the nature of the quantum phase transition manifests at low temperature. Therefore, it is suitable to study low-temperature physics to investigate the behavior of the critical point of the phase transition under momentum dissipation, which is the main focus of this paper.

\para
This paper is organized as follows. In section 2, we introduce the holographic model of Weyl semimetal including axion fields. In section 3, we calculate the dc conductivities of the vector gauge field fluctuations and investigate their behavior with respect to the momentum relaxation strength. Section 4 is devoted to the conclusion and discussion. The Appendix presents the details of the equations of motion and asymptotic expansions.

%%%%%%%%%%%%%%%%%%%%%%%%%%%%%
\section{Holographic Weyl semimetal with momentum relaxation}\label{sec2}
%%%%%%%%%%%%%%%%%%%%%%%%%%%%%

In this section, we begin our setup of the holographic Weyl semimetal with momentum relaxation which is induced by the axion fields. The action for the model reads
\bea\label{action}
\mathcal{S}&=&\int d^5x\sqrt{-g}
  \bigg[\frac{1}{2\kappa^2}\big(R+\frac{12}{L^2}\big)-\frac{1}{4}F^2-\frac{1}{4}\mathcal{F}^2+\frac{\alpha}{3}\epsilon^{abcde}A_a\big(F_{bc}F_{de} +3\mathcal{F}_{bc}\mathcal{F}_{de}\big) \nn\\
  &&-(D_a\Phi)^\ast(D^a\Phi)-V(\Phi)-\frac{1}{2}\sum_{I=1}^{3}(\partial \psi_{I})^2 \bigg]+\mathcal{S}_{GH}+\mathcal{S}_{c.t.} \,,
\eea
where $\kappa^2$, $L$, and $\alpha$ are the gravitational constant, AdS radius, and Chern-Simons coupling, respectively.
According to the holographic dictionary, the vector gauge field $V_a$ corresponds to vector current in the dual field theory with field strength $\mathcal{F}_{ab}=\partial_a V_b-\partial_b V_a$. The axial gauge field $A_a$ corresponds to axial current in the dual field theory with field strength $F_{ab}=\partial_a A_b-\partial_b A_a$. The scalar field $\Phi$ is charged under the axial gauge transformation, and the covariant derivative is $D_a\Phi=(\partial_a-iqA_a)\Phi$. We choose the scalar field potential $ V(\Phi)=m^2\Phi^2+\frac{\lambda}{2} \Phi^4$ with the scalar field mass $m^2=-3$. Therefore, the operator dual to the scalar field has conformal dimension $3$, and its source has conformal dimension 1. Note that the scalar field $\psi_I (I=1,2,3)$ is massless and its total number is equal to the spatial dimension of the dual system. $\mathcal{S}_{GH}$ is the Gibbons-Hawking boundary term, and $\mathcal{S}_{c.t.}$ is the counterterm to demand that the physical observable is finite. Without loss of generality, we will focus on the cases of $q=1$ and $\lambda=1/10$ in the following.

\para
The finite-temperature ansatz for the background fields reads
\bea\label{ansatz}
ds^2=-udt^2+\frac{dr^2}{u}+f(dx^2+dy^2)+hdz^2,\,\,  A=A_z dz,\,\,  \Phi=\phi(r),\,\, \psi_I=\beta_{Ij} x^j \,,
\eea
where the fields $u, f, h, A_z$, and $\phi$ are functions of the radial coordinate $r$. The corresponding equations of motion can be found in the Appendix. Near the UV boundary, $r\to \infty$, we demand that the background geometry is asymptotically to $\text{AdS}_5$ with $u,f,h\sim r^2$. The asymptotic behavior for the axial gauge field and the scalar field reads
\bea
A_z=b+\cdots,\,\,\, \phi=\frac{M}{b}+\cdots ,
\eea
where $M$ and $b$ correspond to the mass parameter and the time-reversal symmetry-breaking parameter in the field theory, respectively. The scalar fields $\psi_I \, (I=1,2,3)$ depend linearly on the spatial coordinate $(x^j={x, y, z})$, where $\beta_{Ij}=\beta$ is a positive real constant. Similar to the particle physics, the scalar fields $\psi_I$ are often called axions, as they have a shift symmetry. The spatial translational symmetry $x^a\to x^a+\xi^a$ is broken due to the spatially dependent sources of $\psi_I$. More precisely, the axion fields will contribute to the Ward identity of the boundary energy-momentum tensor $\nabla_i\langle T^{ij}\rangle=\nabla^j \psi_I^{(0)}\langle O_I \rangle $, which indicates the nonconservation of boundary momentum. Therefore, the axion fields give us a simple holographic approach to dissipate the momentum in the dual field theory, where $\beta$ represents the strength of momentum dissipation.

\subsection{Holographic Weyl semimetal without axion fields }
\para
For $\beta=0 $, the translational symmetry is recovered, and the axion fields drop out of the equations of motion. The holographic Weyl semimetal has been studied in this case \cite{Landsteiner:2015pdh,Landsteiner:2015lsa}, which we will review briefly in this subsection. We will summarize the zero-temperature as well as the finite-temperature physics and discuss how to probe the critical point of the phase transition at low temperature.

\para
At zero temperature, the dual field theory preserves the Lorentz invariance in the $(t,x,y)$ direction, which corresponds to $u=f$. We have only one controllable dimensionless parameter: $M/b$. There exist three different kinds of IR solutions, which correspond to different value of $M/b$, (I){\em the Weyl semimetal phase} for $M/b<(M/b)_c$, (II){\em the Lifshitz critical point} for $M/b=(M/b)_c=0.744$, and (III) {\em the topological trivial phase} for $M/b>(M/b)_c$. At zero temperature, the critical point $(M/b)_c$ is uniquely determined by the Lifshitz critical point. The near-horizon value $A_z(0)$ is nonzero in the Weyl semimetal phase, while it always vanishes in the topological trivial phase. By tuning the parameter $M/b$, the system undergoes a topological quantum phase transition from the Weyl semimetal phase to a topological trivial phase. The order parameter is anomalous Hall conductivity, which is proportional to the near-horizon value of $A_z$:
\bea
\sigma_{\text{AHE}}\propto A_z(0).
\eea

\begin{figure}[!ptb]
\begin{center}
\includegraphics[width=0.6\textwidth]{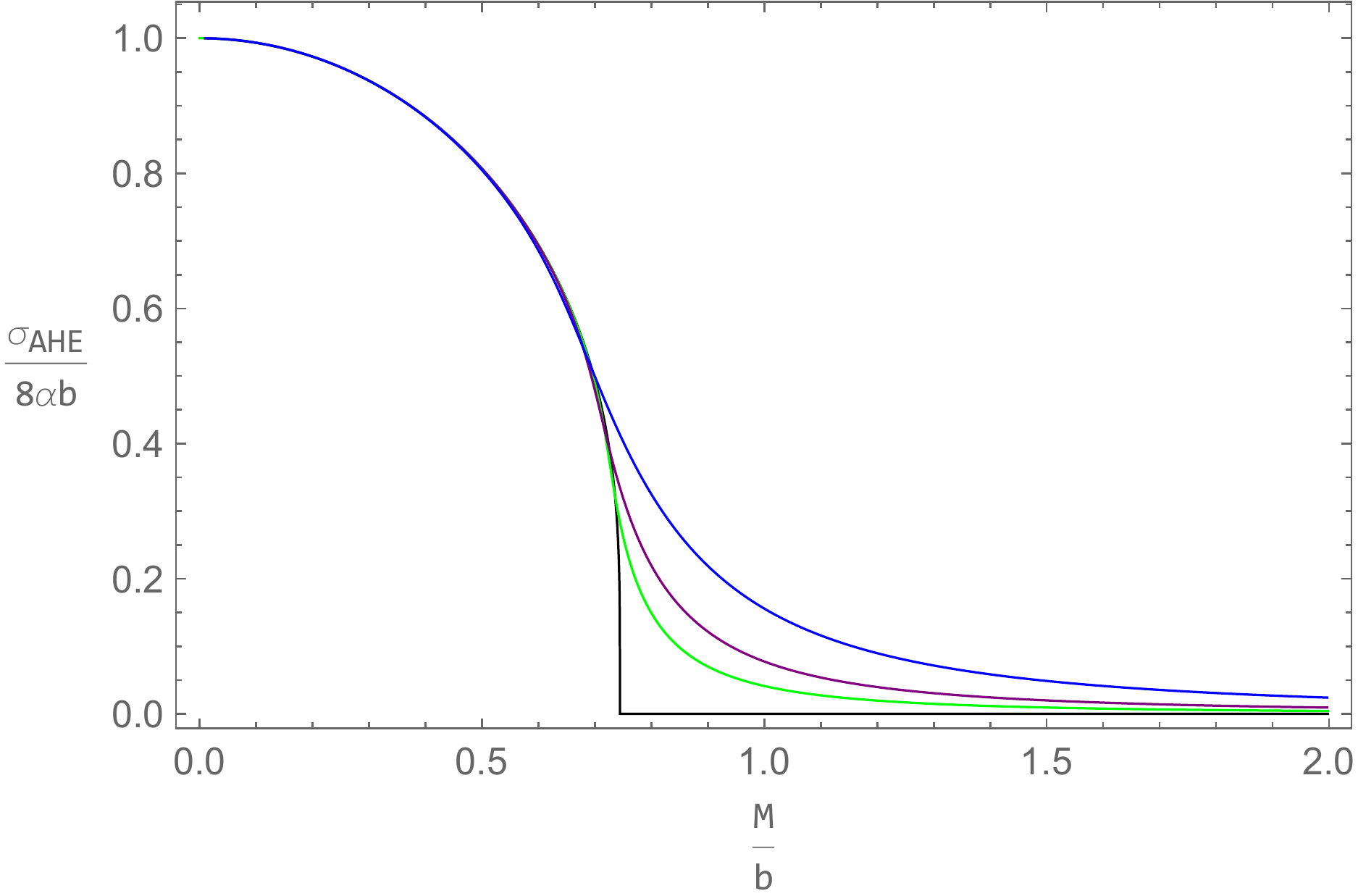}
\end{center}
\vspace{-0.6cm}
\caption{The anomalous Hall conductivity as a function of the $M/b$ without momentum relaxation for different temperatures \cite{Landsteiner:2015pdh}. The black line is for zero temperature, and the colored lines are for finite temperature with $T/b=0.05$ (blue), $0.03$ (purple), and $0.02$ (green), respectively. There is a sharp quantum phase transition at zero temperature which becomes a crossover at finite temperature.}
\label{fig:ahe0}
\end{figure}

\para
At finite temperature, the background solutions admit a regular expansion near the black hole horizon $r=r_h$ with $u(r_h)=0$. We have two dimensionless parameters: $M/b$ and $T/b$. At finite and low temperature, the sharp quantum phase transition becomes a crossover due to the thermal fluctuations, where the anomalous Hall conductivity remains a very small value in the topological trivial phase. Figure \ref{fig:ahe0} shows the anomalous Hall conductivity as a function of the $M/b$ at different temperatures for the holographic Weyl semimetal without momentum relaxation.

%%%%%%%%%%%%%%%%%%%%%%%%%%%%
%%%%%%%%%%%%%%%%%%%%%%%%%%%%%

\para
As the temperature is decreased, the anomalous Hall conductivity approaches that of the ground state, which can be used to probe the location of the critical point of the quantum phase transition. At zero temperature, the critical point is equal to the point with divergent $\vert \frac{\partial\sigma_{\text{AHE} }}{\partial(M/b)} \vert$. At finite temperature, we locate the position of the critical point as the point with maximum $\vert \frac{\partial\sigma_{\text{AHE} }}{\partial(M/b)} \vert$. For example, the critical value obtained at $T/b=0.02$ is $0.722$ with a relative error within $3\% $ for the holographic Weyl semimetal without momentum dissipation. Therefore, the probe of the critical point is accurate at low temperature, and we will use this method to determine the critical point of phase transition in momentum relaxed holographic Weyl semimetal.

%%%%%%%%%%%%%%%%%%%%%%%%%%%%%
\subsection{Holographic Weyl semimetal with axion fields}
%%%%%%%%%%%%%%%%%%%%%%%%%%%%%

\para
In the presence of axion fields with $\beta\neq0$, the holographic Weyl semimetal is supposed to still exhibit a quantum phase transition between the Weyl semimetal phase and the topological trivial phase. At zero temperature, the theory is characterized by two dimensionless parameters: $M/b$ and $\beta/b$.  Different from the minimal model \cite{Landsteiner:2015pdh}, the zero-temperature solutions have $u\neq f$, which can also be observed from the background equations of motion in the Appendix. Therefore, it is difficult to find the ground state of the holographic Weyl semimetal in the presence of axion fields, and we will leave it for further work.

\para
At finite temperature, we have three dimensionless parameters: $M/b$, $T/b$, and $\beta/b$. The asymptotic expansions for the background fields change slightly compared with the minimal model; see the Appendix for more details. We focus on the low-temperature physics and fix the temperature of the system to be $T/b=0.02$. Therefore, using the shooting method, we can obtain a series of numerical solutions of the background equations of motion which depends on the remaining two dimensionless parameters ($M/b$ and $\beta/b$). In the next section, we will study the effects of momentum relaxation on the order parameter and various dc conductivities.

%%%%%%%%%%%%%%%%%%%%%%%%%%%%%
\section{Momentum relaxation effects on the phase transition}\label{sec3}
%%%%%%%%%%%%%%%%%%%%%%%%%%%%%
\para
To explore the effects of momentum relaxation, we study the conductivities, i.e., the response of the background system under the gauge field fluctuations. In the following, we will obtain the phase diagram of the holographic Weyl semimetal from the anomalous Hall conductivity. We will compute the longitudinal and transverse dc conductivities. We will also study the behavior of dc resistivity as a function of temperature in the two phases.

\para
The conductivities of the dual field system are related to the retarded current-current correlation via the Kubo formula:
\bea
\sigma_{ij}=\lim_{\omega\to 0}\frac{1}{i\omega}\langle J_i J_j\rangle_R(\omega, \mathbf{k}=0) \,.
\eea
In holography, the retarded Green's functions can be obtained from the dual gauge fields fluctuations above the background solutions, where the infalling boundary conditions are imposed at the black hole horizon.

\para
We turn on the vector gauge field fluctuations along the spatial directions
\bea
\delta V_x=v_x(r)e^{-i\omega t},\,\,  \delta V_y=v_y(r)e^{-i\omega t},\,\,  \delta V_z=v_z(r)e^{-i\omega t} \,,
\eea
Note that, since the vector field perturbations decouple from that of the axion fields, we do not need to consider the fluctuations of the axion fields like Ref. \cite{Andrade:2013gsa}. Generally, the axion fields affect the physical system in two aspects. First, they alter the background solution and its thermodynamics. Second, they cause the momentum relaxation by directly coupling to the perturbation fields. For the model we studied here, the absence of axion fields in the vector field fluctuations indicates that their effects on the transports arise from their effects on the equilibrium solution.

\para
Plugging the above ansatz into the vector field equations, we find
\bea
v_z''+\bigg(\frac{u'}{u}+\frac{f'}{f}-\frac{h'}{2h}\bigg)v_z'+\frac{\omega^2}{u^2}v_z&=&0  \,,\\
v_\pm''+\bigg(\frac{u'}{u}+\frac{h'}{2h}\bigg)v_\pm'+\frac{\omega^2}{u^2}v_\pm \pm 8\alpha\omega\frac{A_z'}{u\sqrt{h}}v_\pm&=&0  \,,
\eea
where we have defined $v_\pm=v_x\pm i v_y$ to get the last equation.

\para
The full frequency conductivities can be obtained numerically by solving the above equations with an ingoing boundary condition at the horizon. However, since we are interested in only the dc conductivities, we will alternatively use the near-far matching method to obtain the desired results, following Ref. \cite{Landsteiner:2015pdh}. This method treats the above equations semianalytically, and its final results can be expressed in terms of data on the black hole horizon $r_h$. The dc conductivities $\sigma_{xx}, \sigma_{yy}$, and $\sigma_{xy}$ can be computed along this procedure as
\bea\label{eq:ahe}
\sigma_T=\sigma_{xx}=\sigma_{yy}=\frac{G_{+}+G_{-}}{2i\omega}=\sqrt{h(r_h)},\,\,  \sigma_{xy}=\frac{ G_{+}-G_{-} }{2\omega}=8\alpha \big( b-A_z(r_h) \big) \,,
\eea
where $G_{\pm}=\omega \big( \pm 8\alpha(b-A_z(r_h))+i \sqrt{h(r_h)} \big)$ are the Green functions of $v_\pm$. Using the same method, the longitudinal conductivity $\sigma_{zz}$ is given by
\bea
\sigma_{zz}=\frac{G_{zz}}{i\omega}=\frac{f(r_h)}{\sqrt{h(r_h)}} \,.
\eea
%%%%%%%%%%%%%%%%%%%%%%%%%%%%%
\subsection{Phase diagram}\label{sec:pd}
%%%%%%%%%%%%%%%%%%%%%%%%%%%%%
\para
The phase transition is characterized by the anomalous Hall conductivity, which can be expressed as
\bea\label{ahe}
\sigma_{\text{AHE} }=8\alpha b-\sigma_{xy}=8\alpha A_z(r_h) \,.
\eea

\begin{figure}[!ptb]
\begin{center}
\includegraphics[width=0.45\textwidth]{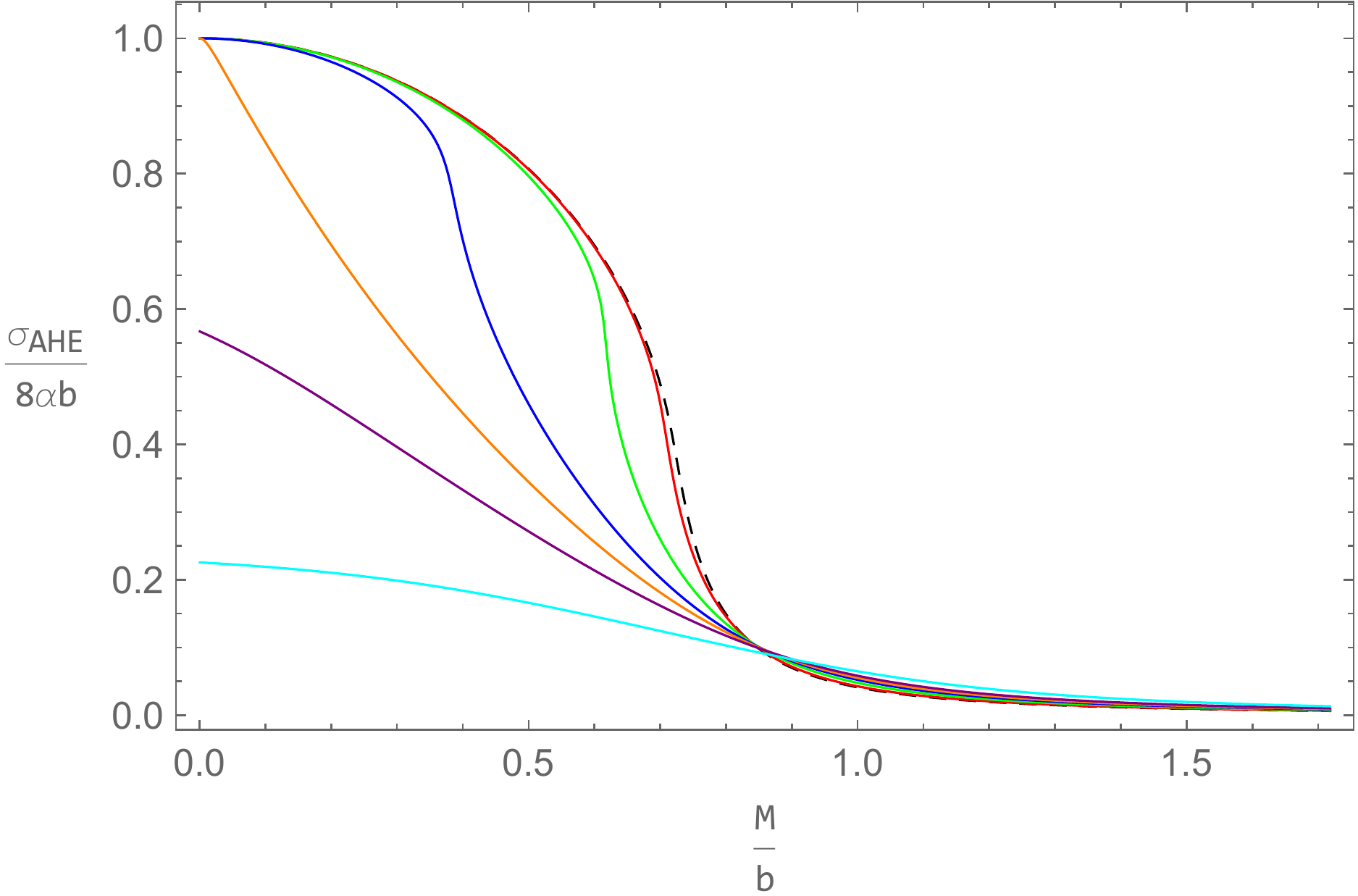}
\includegraphics[width=0.45\textwidth]{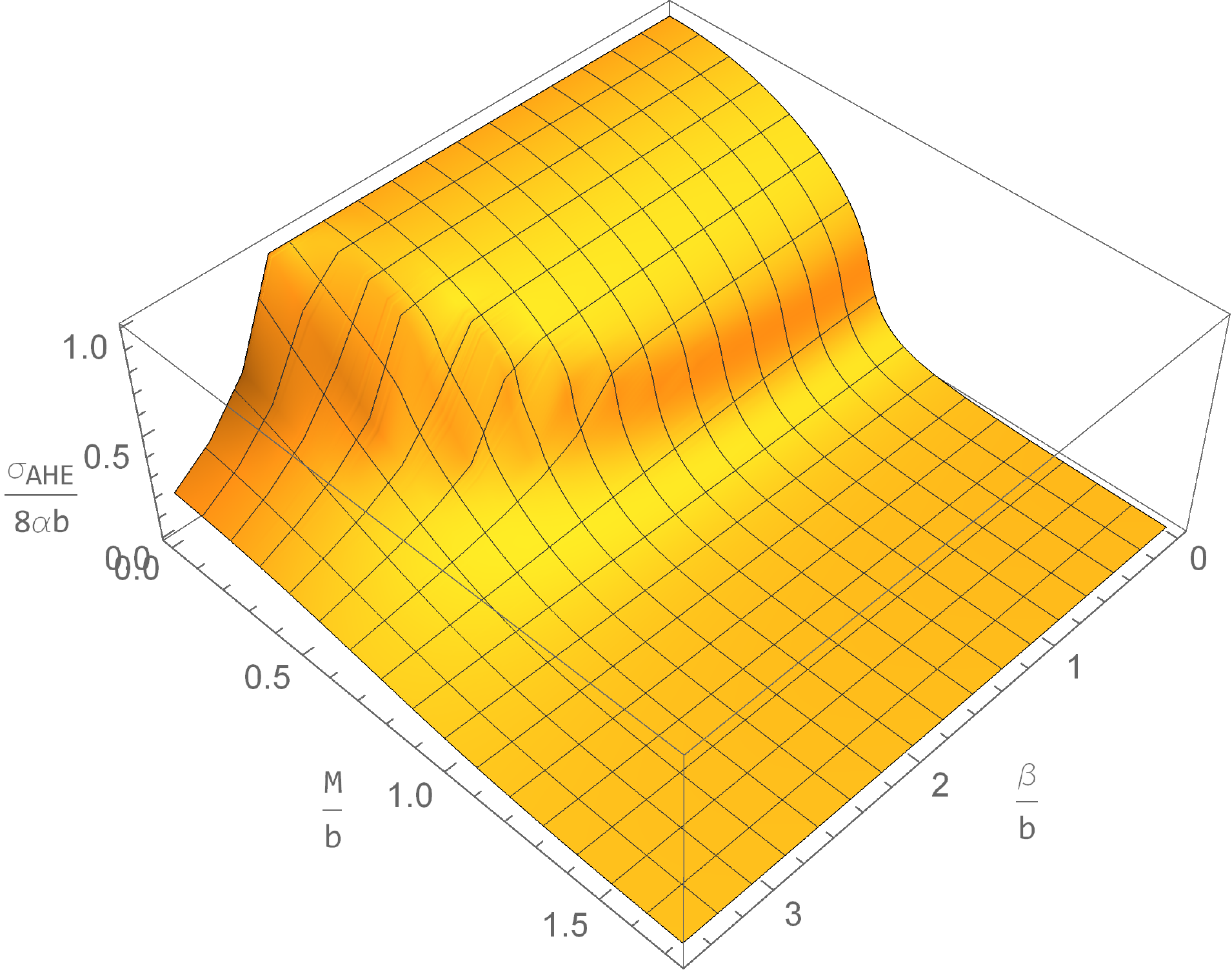}
\end{center}
\vspace{-0.6cm}
\caption{Left: the normalized anomalous Hall conductivity as a function of the $M/b$ for different $\beta/b$ at temperature $T/b=0.02$. The black dashed curve is the anomalous Hall conductivities without momentum dissipation, while the colored curves are for $\beta/b=1$ (red), $2$ (green), $2.5$ (blue), $2.75$ (orange), $3$ (purple), and $3.5$ (cyan), respectively. Right: the 3D version of the anomalous Hall conductivity as functions of $M/b$ and $\beta/b$, where the value of $\beta/b$ ranges from $0$ to $3.5$ with an interval $1/4$.}
\label{fig:ahe}
\end{figure}

\para
In Fig. \ref{fig:ahe}, we plot the anomalous Hall conductivity as a function of $M/b$ for different $\beta/b$ at temperature $T/b=0.02$. For a fixed value of $\beta/b$, the figure shows that, as we increase $M/b$, the anomalous Hall conductivity decreases monotonically from the Weyl semimetal phase to a very small value in the topological trivial phase. By comparing with the results without momentum relaxation (black dashed curve), we find that the momentum relaxation affects the phase transition in an interesting way. For a small value of momentum relaxation strength (i.e., for $\beta/b<1$), the anomalous Hall conductivity remains almost unchanged. As we increase the value of $\beta/b$ further, the anomalous Hall conductivity changes dramatically, and its value decreases rapidly in the region $M/b<0.744$ (i.e., the original Weyl semimetal phase with $\beta/b=0$). As the topological Weyl semimetal phase is characterized by a nontrivial anomalous Hall conductivity, the behavior of anomalous Hall conductivity may indicate that the region of Weyl semimetal phase narrows and finally disappears with the increase of $\beta/b$. \footnote{At $M/b=0.01$, the numerical results of AHE are less than $1$ for $\beta/b>2.75$. This seems inconsistent with the analytical results at $M/b=0$, which we will explain in the next subsection.}
%%%%%%%%%%%%%%%%%%%%%%%%%%%%%
%%%%%%%%%%%%%%%%%%%%%%%%%%%%%

\begin{figure}[!ptb]
\begin{center}
\includegraphics[width=0.6\textwidth]{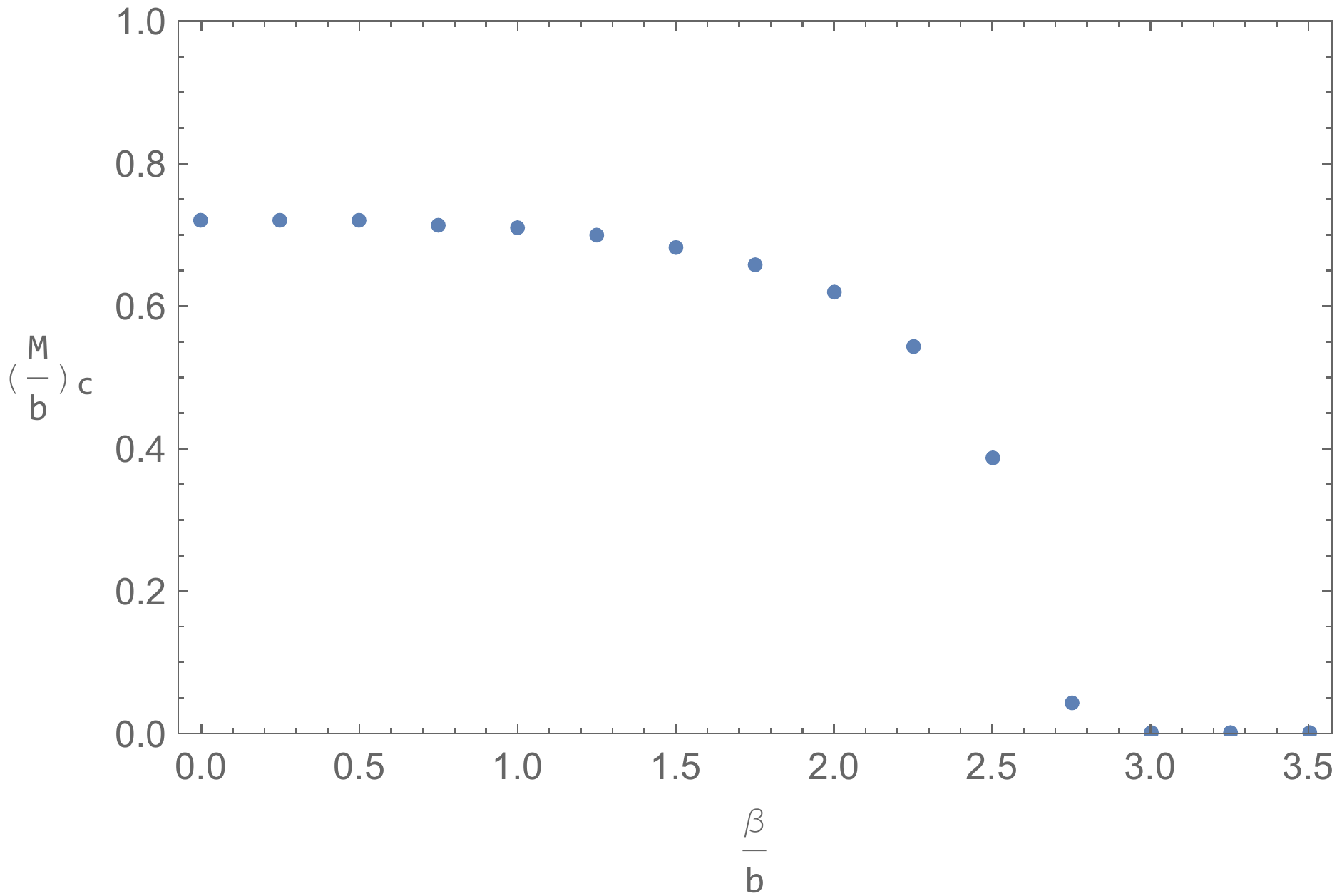}
\end{center}
\vspace{-0.6cm}
\caption{The critical point of the phase transition as a function of $\beta/b$ at temperature $T/b=0.02$.}
\label{fig:critical}
\end{figure}

\para
In order to characterize more specifically the effects of momentum relaxation on the order parameter, we will study the behavior of the critical point of the phase transition with respect to $\beta/b$. The critical point can be obtained from the anomalous Hall conductivity, which is equivalent to the point with maximum $\vert \frac{\partial\sigma_{\text{AHE} } }{\partial(M/b)} \vert$. We show our main results in Fig. \ref{fig:critical}, which gives the behavior of critical point $(M/b)_c$ as a function of $\beta/b$ at temperature $T/b=0.02$. As we increase $\beta/b$, the value of the critical point decrease monotonically. There exists a critical $(\beta/b)_c$, above which the value of the critical point becomes zero. The value of the critical point decreases very slowly for $\beta/b<2$, while it gets smaller rapidly for $2<\beta/b<(\beta/b)_c$. \footnote{For $2.5< \beta/b<(\beta/b)_c$, the $\vert \frac{\partial\sigma_{\text{AHE} } }{\partial(M/b)} \vert$ does not show a sharp peak, which means that critical value $(M/b)_c$ in this region may have a relatively large error.} This indicates that the momentum relaxation can reduce and even destroy the Weyl semimetal phase, which is the main findings of this paper. This is consistent with the field theory predictions, and we will give a simple explanation as follows. From the dual point of view, we fix the distance of Weyl points in the momentum space to be 1. The case of $\beta/b=0$, i.e., for a system without momentum relaxation, corresponds to the width of Brillouin zone $k_L\to \infty$. As we increase the momentum relaxation strength $\beta/b$, the value of $k_L$ decreases. There exists a critical $\beta/b$ to make $k_L=1$ where the two Weyl points meet and annihilate each other due to the periodicity of the Brillouin zone. This picture explains the observed disappearance of the Weyl semimeatl phase as $\beta/b$ is increased.

%%%%%%%%%%%%%%%%%%%%%%%%%%%%%
%%%%%%%%%%%%%%%%%%%%%%%%%%%%%
%%%%%%%%%%%%%%%%%%%%%%%%%%%%%
\subsubsection{ Anomalous Hall conductivity at $M/b=0$ }
%%%%%%%%%%%%%%%%%%%%%%%%%%%%%
\para
In this subsection, we will analyze the two possible solutions of the momentum relaxed system in the $M/b\to 0$ limit and then explain the apparent conflict mentioned in the footnote of the above subsection. In the $M/b\to 0$ limit, the background geometry has a simple analytical solution \cite{Andrade:2013gsa}, which reads
\bea\label{sch}
u=r^2-\frac{r_h^4}{r^2}+\frac{\beta^2}{4}\Big( -1+\frac{r_h^2}{r^2} \Big) ,\,\,  f=h=r^2,\,\,  A_z=b,\,\,  \phi=0,\,\,  \psi_I=\beta x^I \,,
\eea
From Eq. (\ref{ahe}), the normalized anomalous Hall conductivity is $\frac{\sigma_{\text{AHE} }}{8\alpha b}|_{\frac{M}{b}=0}=1$, which is independent of $\beta/b$. In addition to the analytical solution, we can find a spontaneous symmetry-breaking-type solution following the analysis in Ref. \cite{Horowitz:2009ij}. At zero temperature $r_h=\frac{\beta}{2\sqrt{2} }$, the near-horizon limit of Eq. (\ref{sch}) is $\text{AdS}_2\times \mathbb{R}^3$. By analyzing the linearized equation of motion for $\phi$, we find that its effective mass at the extremal geometry becomes $m_{eff}^2=\frac{m^2}{4}+\frac{2b^2q^2}{\beta^2}$. Therefore, the zero-temperature background is unstable if $m_{eff}^2$ is below the Breitenlohner-Freedman (BF) bound of the $\text{AdS}_2$: $m_{BF}^2=-1/4$. At zero temperature, the condition for instability is $\beta/b>\frac{2\sqrt{2}q}{\sqrt{1-m^2}}$. For the particular parameters we studied in this paper, the new branch of solution becomes more pronounced if $\beta/b>\sqrt{2}$. At finite temperature with $T/b$ fixed, there exists a critical $(\beta/b)_{n} $ above which the new solution appears.\footnote{We conjecture that this critical $(\beta/b)_n$ is equal to the $(\beta/b)_{c} $ shown in Fig. \ref{fig:critical}.}

\para
From the above analysis, we know that there exist two solutions for $M/b=0$ at temperature $T/b=0.02$. Therefore, the apparent inconsistency can be understand as follows. As $M/b\to 0$, the numerical solutions in Fig. \ref{fig:ahe} approach the spontaneous symmetry-breaking-type solution more easily when the value of $\beta/b$ is larger than $2.75$. A detailed analysis of the various phases near this region is beyond the scope of this paper and needs more further work.

%%%%%%%%%%%%%%%%%%%%%%%%%%%%%
\subsection{dc conductivities and resistivities}
%%%%%%%%%%%%%%%%%%%%%%%%%%%%%

\begin{figure}[!ptb]
\begin{center}
\includegraphics[width=0.6\textwidth]{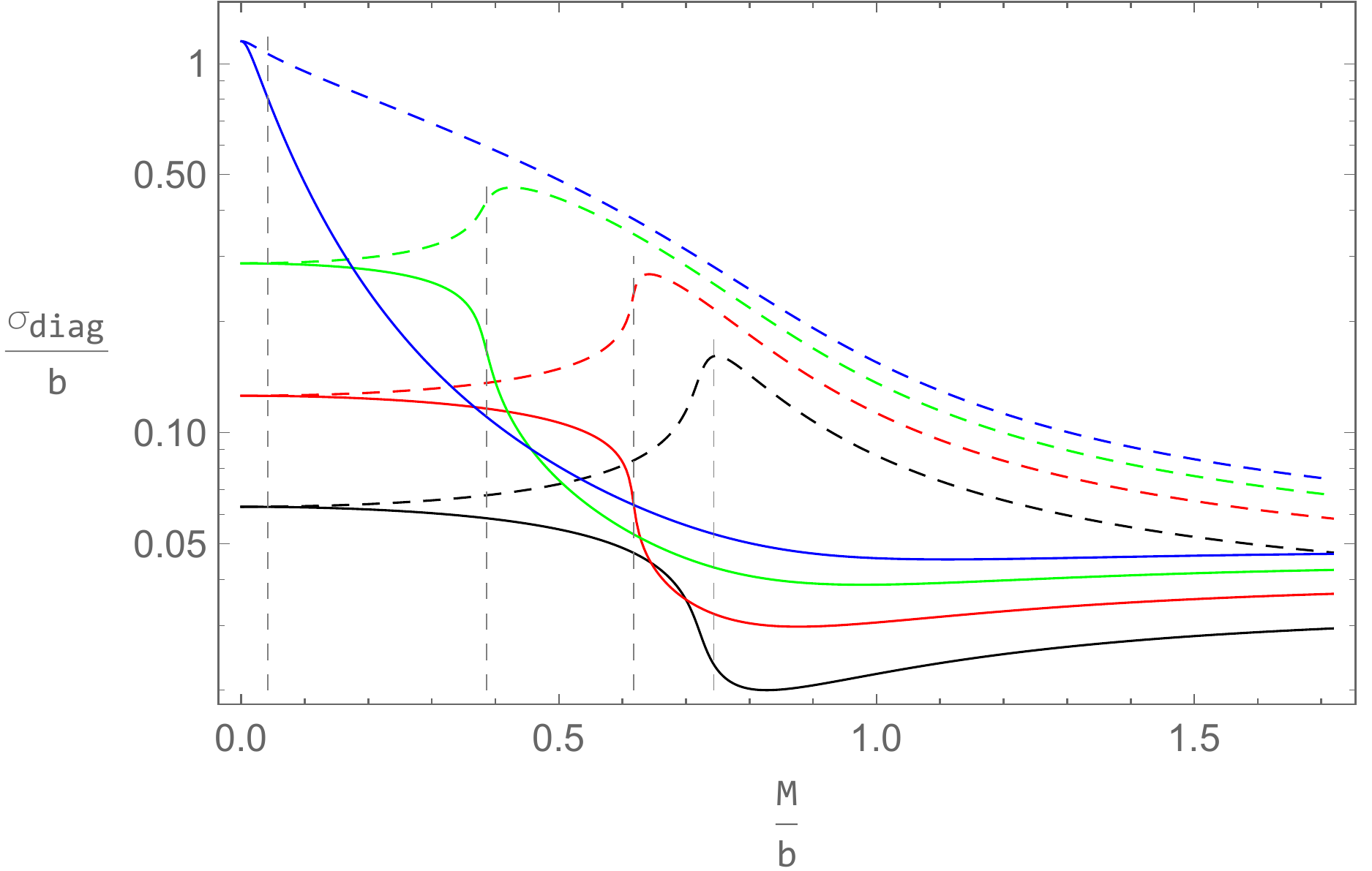}
\end{center}
\vspace{-0.6cm}
\caption{The linear-log plot of the transverse (dashed lines) and longitudinal (solid lines) conductivities as a function of the $M/b$ for different $\beta/b$ at temperature $T/b=0.02$. The black lines are the conductivities of the system without momentum relaxation, while the colored curves are for $\beta/b=2$ (red), $2.5$ (green), and 2.75 (blue), respectively. The dashed gray lines are the positions of the critical points of the phase transition.}
\label{fig:dcTL}
\end{figure}

\para
Apart from the anomalous Hall conductivity, it is interesting to study the behavior of the diagonal conductivities as a function of $M/b$ for different $\beta/b$ at $T/b=0.02$. Figure \ref{fig:dcTL} shows that the transverse (longitudinal) conductivities produce a peak (minimum) at an intermediate value of $M/b$, where the location of the peak (minimum) decreases as $\beta/b$ is increased. By comparing the location of the peak (minimum) with the critical value of the phase transition (vertical lines), we find that they both have a similar monotonically decreasing behavior with the increase of $\beta/b$. This supports the results of the shrink and disappearance of the Weyl semimetal phase under the momentum dissipation observed from the behavior of anomalous Hall conductivity. As $M/b\to 0$, we find that the transverse and longitudinal conductivities have the same value at fixed $\beta/b$ and the value increases as $\beta/b$ is increased. For large $M/b$, the diagonal conductivities approach constant values, and their values increase slightly as $\beta/b$ is increased.
%%%%%%%%%%%%%%%%%%%%%%%%%%%%%
%%%%%%%%%%%%%%%%%%%%%%%%%%%%%

\begin{figure}[!ptb]
\begin{center}
\includegraphics[width=0.42\textwidth]{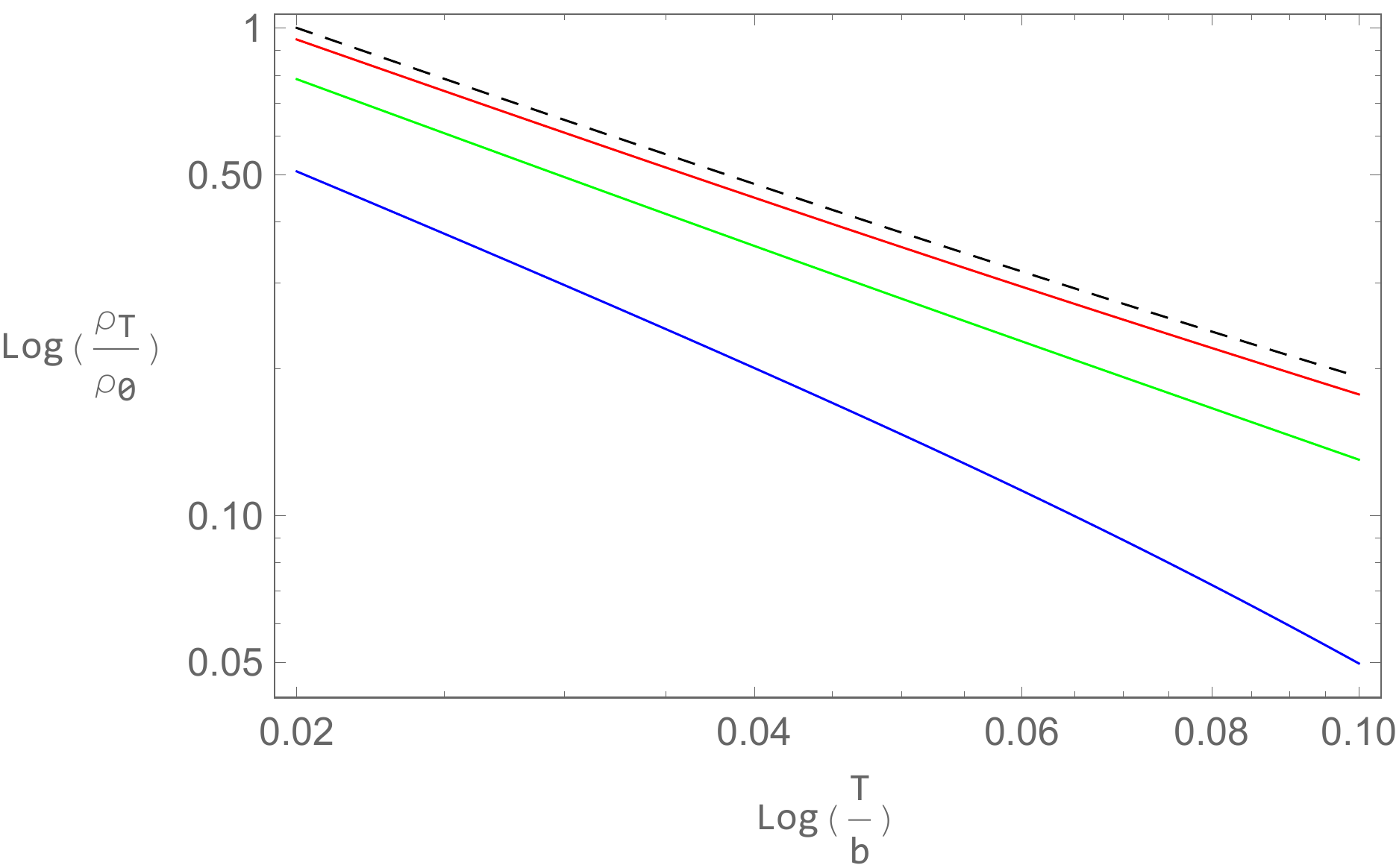}
\includegraphics[width=0.42\textwidth]{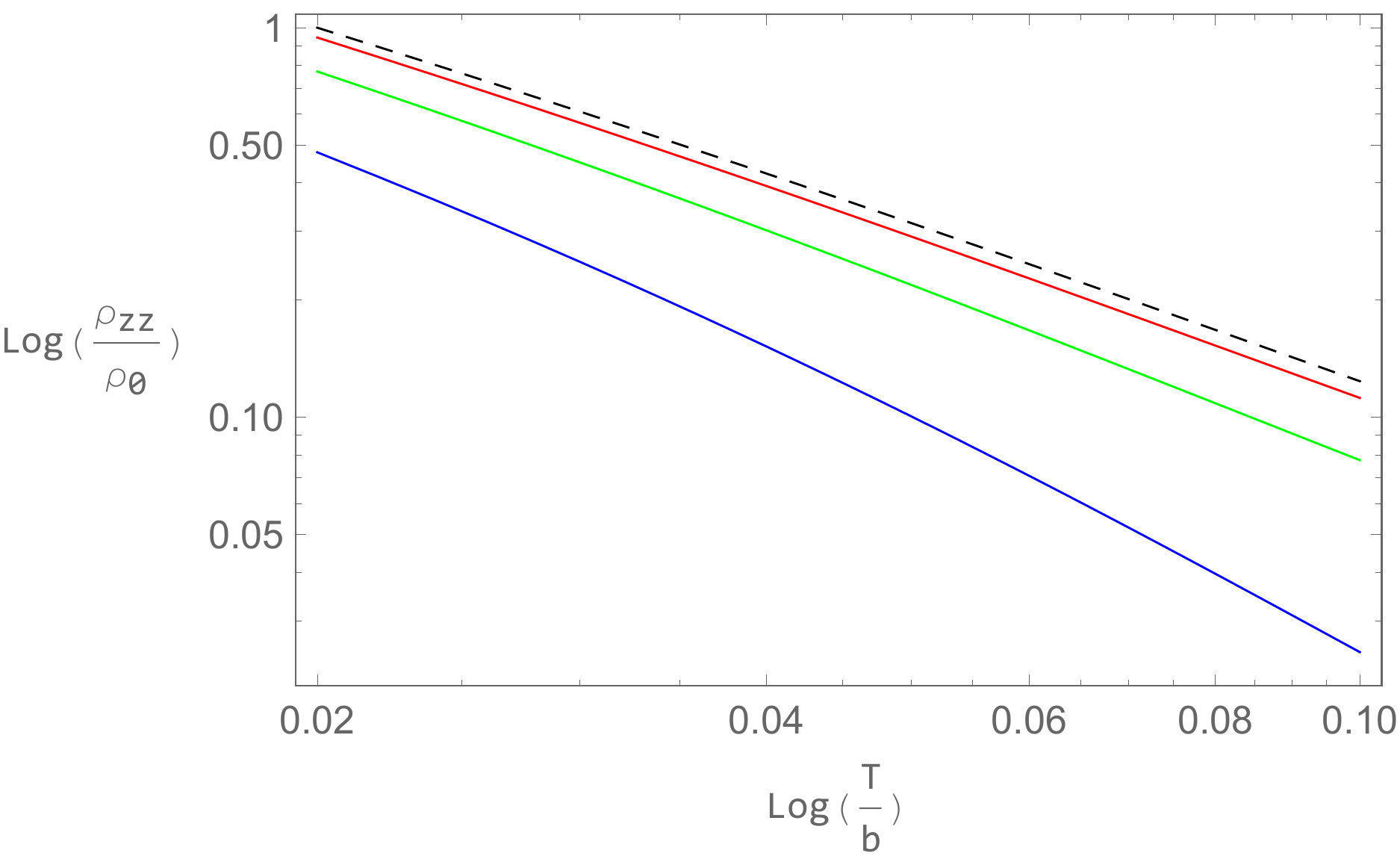}
\includegraphics[width=0.42\textwidth]{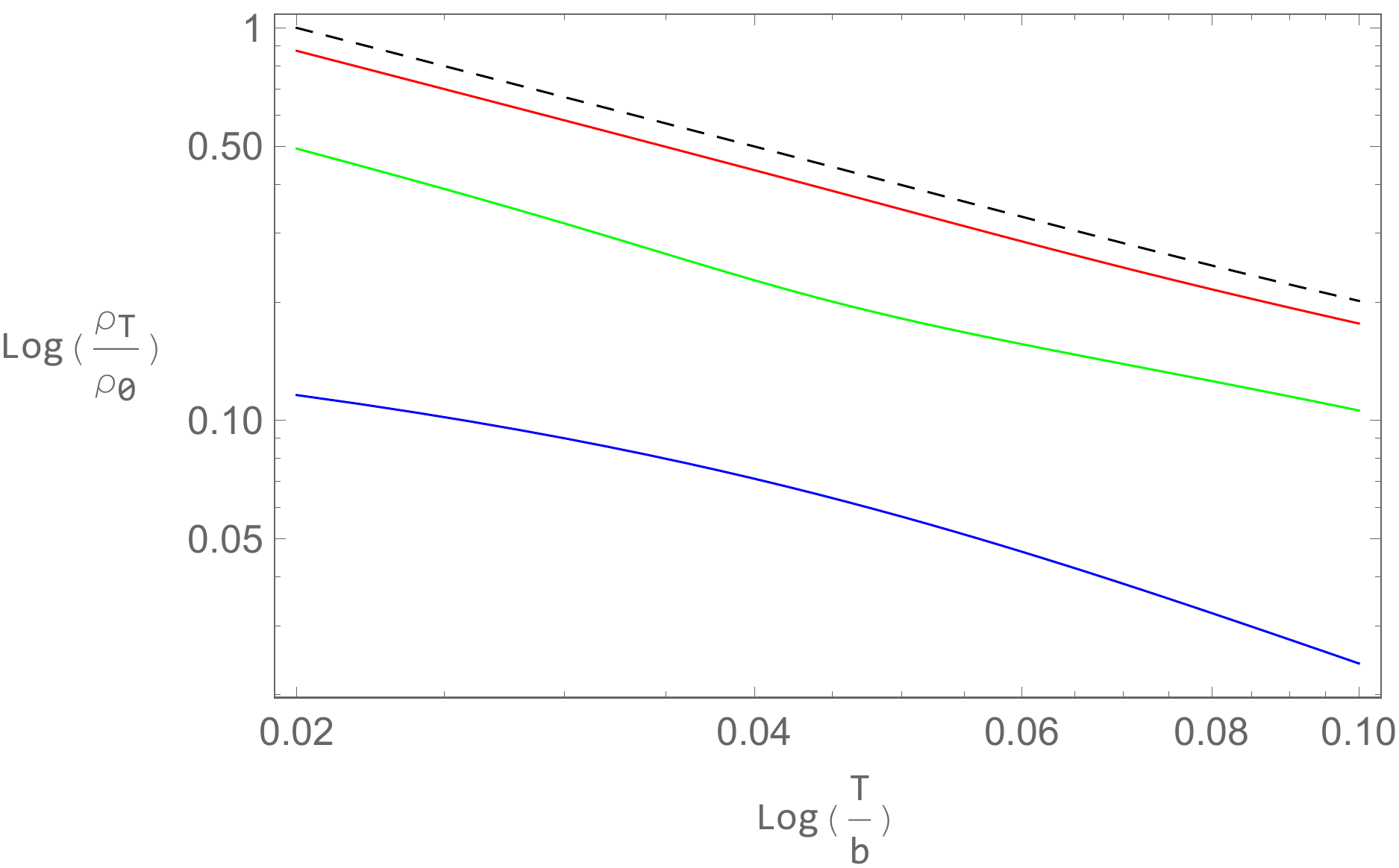}
\includegraphics[width=0.42\textwidth]{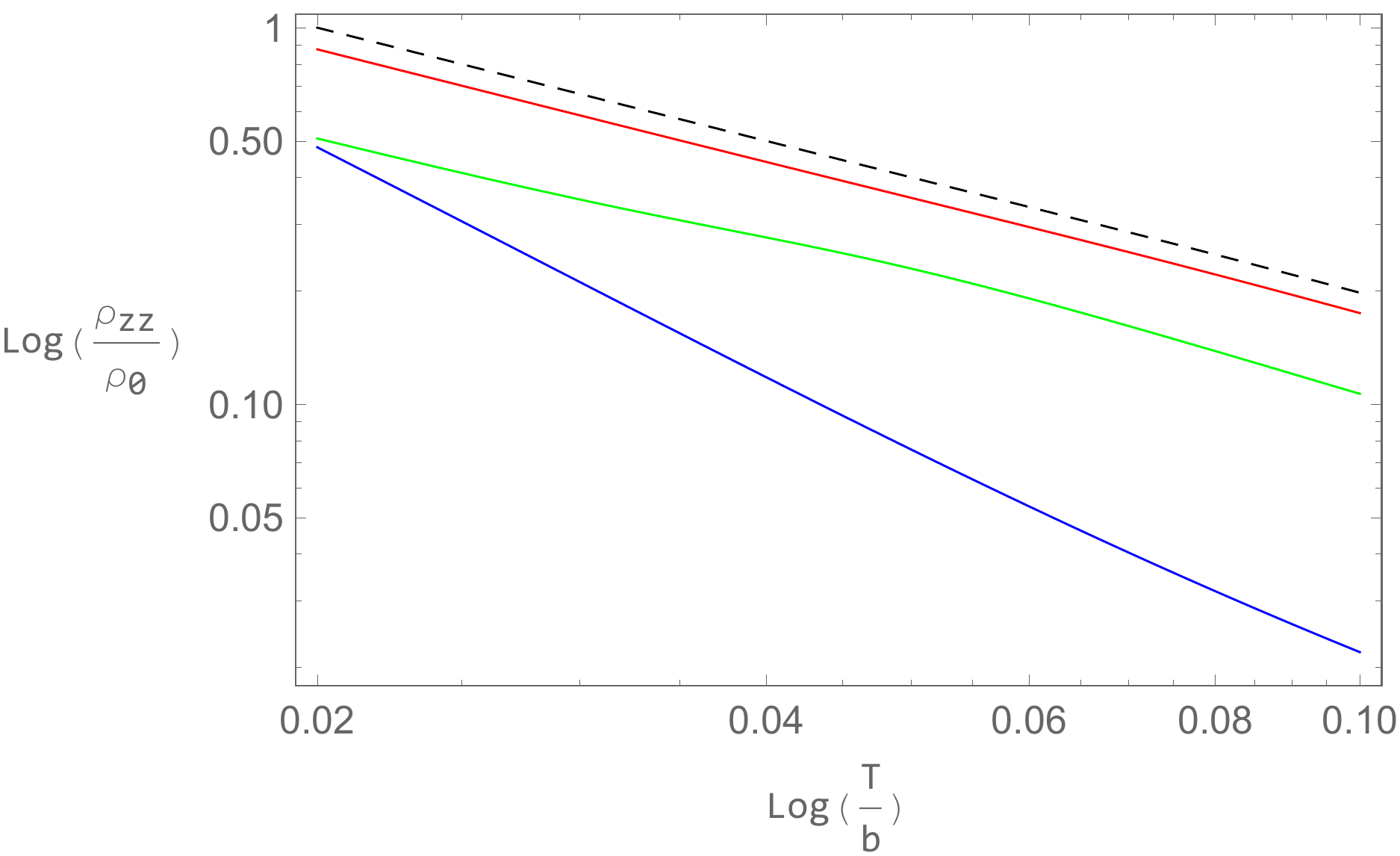}
\end{center}
\vspace{-0.6cm}
\caption{Log-log plot of the transverse(left) and longitudinal(right) dc resistivities as a function of the temperature for different $\beta/b$, where the top two panels are for the topological trivial phase with $M/b=1.2$ and the bottom two panels are for the Weyl semimetal phase with $M/b=0.45$. The black dashed line is the dc resistivity of the system without momentum relaxation, and the colored curves are for $\beta/b=1$ (red), $2$ (green), and $3$ (blue), respectively, in each panel. The resistivity $\rho_0$ is for the normalization of the dc resistivity in each panel.}
\label{fig:mi}
\end{figure}

\para
Figure \ref{fig:mi} shows the dc resistivity $\rho=1/\sigma$ as a function of the temperature in the topological trivial phase and the Weyl semimetal phase. At low temperature, the dc resistivity decreases as a function of the temperature for fixed $\beta/b$ in the two phases. For $\beta/b=0$, the dc resistivities (black dashed curve) in both phases behave as $\rho_{T/zz}\sim T^{-1}$, which corresponds to the linear dependence of the conductivities $\sigma_{T/zz}\sim\omega  (\omega\to 0,\, T=0) $ in the ground state \cite{Grignani:2016wyz,Landsteiner:2015pdh}. As we increase $\beta/b$, the behavior of the dc resistivities changes gradually, and the linear $T^{-1}$ dependence is inapplicable. In the topological trivial phase, the dc resistivities have a power law dependence as $\rho\sim T^{-1-\delta}$, where the value of $\delta$ depends on $\beta/b$. This power law scaling reveals a possible emergent symmetry of the zero-temperature ground state. In contrast, we do not find a simple scaling behavior for dc resistivities at $M/b=0.45$ with nonzero momentum relaxation strength.

%%%%%%%%%%%%%%%%%%%%%%%%%%%%%
%%%%%%%%%%%%%%%%%%%%%%%%%
\section{Conclusion and discussion}\label{sec4}
%%%%%%%%%%%%%%%%%%%%%%%%%
\para
In this work, we have studied the momentum relaxation effect in the holographic Weyl semimetal with a topological quantum phase transition. The momentum relaxation is induced by the axion fields in holography which break translational symmetry along spatial directions. The order parameter of the phase transition is the anomalous Hall conductivity. By tuning the momentum dissipation strength, we obtain the behavior of the anomalous Hall conductivity across the phase transition at finite temperature. At finite and low temperature, the critical value of the phase transition can be obtained from the anomalous Hall conductivity. We found that it decreases as the strength of momentum dissipation is increased up to a special value, above which the critical value goes to zero. This indicates that the momentum relaxation can lead to the shrink and disappearance of the Weyl semimetal phase, which is consistent with the predictions of the weakly coupled field theory.

\para
We have also studied the behavior of the transverse and longitudinal conductivities for different momentum relaxation strengths. We found that the maximal (minimal) value of the transverse (longitudinal) conductivity approaches zero when we increase the momentum relaxation strength, which supports the results of the shrink and disappearance of the Weyl semimetal phase under the momentum relaxation. Finally, we have studied the temperature dependence of the dc resistivity for different momentum relaxation strengths in the two phases, where a power law scaling of dc resistivity is observed in the topological trivial phase.

\para
The momentum relaxation affects the holographic Weyl semimetal system in an interesting way, and there are several further questions worthy to explore. First, the shrink and disappearance of the Weyl semimetal phase under the momentum relaxation is observed from the behavior of the critical point of the phase transition. However, the definition of the critical point at finite temperature is not exact compared with the result of the zero-temperature ground state. Therefore, it is important to explore the zero-temperature physics of the translational invariant broken Weyl semimetal in order to get more evidence and an explanation of the phenomenon we found in this paper. Second, as the momentum relaxation is induced by the massless axion fields, the results we found in this paper may depend on the particular translational symmetry-broken mechanics we used. It would be interesting to apply other translational invariant breaking mechanics, like massive gravity \cite{Vegh:2013sk,Davison:2013jba}, to test the universality of our results.

%%%%%%%%%%%%%%%%%%%%%
\vspace{.8cm}
\subsection*{Acknowledgments}

I thank Yan Liu for his suggestions on the project and helpful guidance throughout the work. I thank Hong-Da Lyu, Yan Liu, and Xin-Meng Wu for reading a preliminary version of the manuscript and providing useful comments and suggestions. I thank Zhi-Hong Li, Jie Jiang, Qi-Rong Jiao, Han-Qing Shi, and Hai-Qing Zhang for useful discussions. This work is supported by the National Natural Science Foundation of China Grant No. 11875083.

\vspace{.3 cm}
%%%%%%%%%%%%%%%%%%%%%%%%%%%%%%%%
\appendix
\section{Equations of motion and asymptotic expansions}{\label{app:a}}

We set $2\kappa^2=L=1$ in this paper. The bulk equations of motion corresponding to the ansatz (\ref{ansatz} ) are
\bea
\frac{u''}{u}-\frac{f''}{f}+\frac{h'}{2h}\left(\frac{u'}{u}-\frac{f'}{f}\right)-\frac{\beta ^2}{u f}&=&0  \,, \nn\\
\frac{u''}{2u}+\frac{f''}{f}+\frac{u'f'}{uf}-\frac{f'^2}{4f^2}-\frac{6}{u}-\frac{A_z'^2}{4h}+
\frac{\phi^2}{2u}\left(m^2+\frac{\lambda}{2}\phi^2 -\frac{q^2A_z^2}{h}\right)
+\frac{\phi'^2}{2}+\frac{\beta^2}{2uf}-\frac{\beta^2}{4uh}&=&0  \,,\nn \\
\frac{6}{u}-\frac{u'}{2u}\left(\frac{f'}{f}+\frac{h'}{2h}\right)-\frac{f'h'}{2fh}-\frac{f'^2}{4f^2}
+\frac{A_z'^2}{4h}
-\frac{\phi^2}{2u}\left(m^2+\frac{\lambda}{2}\phi^2+\frac{q^2A_z^2}{h}\right)+\frac{\phi'^2}{2}
-\frac{\beta^2}{2uf}-\frac{\beta^2}{4uh}&=&0   \,, \nn \\
A_z''+\left(\frac{u'}{u}+\frac{f'}{f}-\frac{h'}{2 h}\right)A_z'-\frac{2q^2 \phi^2}{u}A_z&=&0  \,, \nn \\
\phi''+\left(\frac{u'}{u}+\frac{f'}{f}+\frac{h'}{2h}\right)\phi'-\left(\frac{q^2 A_z^2}{uh}+\frac{m^2}{u}\right)\phi-\frac{\lambda \phi^3}{u}&=&0  \,,\nn
\eea
where the prime denotes the derivative with respect to $r$. Note that the first equation can be written as $\big(\sqrt{h}(u'f-uf')\big)'=\beta^2\sqrt{h}$, which is different with the minimal model \cite{Landsteiner:2015pdh}. This indicates that the zero-temperature ground state has a new geometrical configuration with $u\neq f$. Our system has the following three scaling symmetries:\footnote{Note that these scaling symmetries exist before setting $\beta_{Ij}=\beta \delta_{Ij}$ as $\beta_{Ij}$ will change according to each scaling symmetry. After that we are free to set $\beta_{Ij}=\beta\delta_{Ij}$ and obtain the above equations of motion.}\\
(I)~~~$(x,y)\to\gamma(x,y),\,\, f\to \gamma^{-2}f\,; $ \\
(II)~~$z\to\gamma z,\,\,  h\to \gamma^{-2}h, \,\,  A_z\to \gamma^{-1} A_z \,; $ \\
(III)~$r\to\gamma r,\,\, (t,x,y,z)\to\gamma^{-1}(t,x,y,z), \,\, (u,f,h,)\to\gamma^{2}(u,f,h), \,\,  A_z\to\gamma A_z \,; $\\
At the AdS boundary, i.e., as $r\to \infty$, the fields behave as
\bea
u&=&r^2-\frac{M^2}{3}-\frac{\beta^2}{4}+\frac{M^4(2+3\lambda)}{18}\frac{\ln r}{r^2}+\frac{u_2}{r^2}+\cdots \,, \nn\\
f&=&r^2-\frac{M^2}{3}+\frac{-3M^2\beta^2+4M^4(2+3\lambda)}{72}\frac{\ln r}{r^2}+\frac{f_2}{r^2}+\cdots \,, \nn\\
h&=&r^2-\frac{M^2}{3}+\frac{-3M^2\beta^2+36q^2M^2b^2+4M^4(2+3\lambda)}{72}\frac{\ln r}{r^2}+\frac{h_2}{r^2}+\cdots\, , \nn\\
A_z&=&b-q^2M^2b\frac{\ln r}{r^2}+\frac{\eta}{r^2}+\cdots \,,  \nn\\
\phi&=&\frac{M}{r}+\frac{3\beta^2M-12q^2Mb^2-4M^3(2+3\lambda)}{24}\frac{\ln r}{r^3}+\frac{O}{r^3}+\cdots \,.  \nn
\eea

Near the black hole horizon, we have the expansions
\bea
u&=&4\pi T(r-r_h)+\Big(-2+\frac{\beta^2}{2f_0}+\frac{\beta^2}{4h_0}+\frac{m^2\phi_0^2}{6}+\frac{q^2Az_0^2\phi_0^2}{2h_0}+\frac{\lambda \phi_0^4}{12} \Big)(r-r_h)^2+\cdots \,, \nn\\
f&=&f_0-\frac{ \beta^2+f_0(-8+\frac{2m^2\phi_0^2}{3}+\frac{\lambda \phi_0^4}{3} )}{4\pi T}(r-r_h)+\cdots \,, \nn\\
h&=&h_0-\frac{ \beta^2+2q^2A_{z0}^2\phi_0^2+h_0(-8+\frac{2m^2\phi_0^2}{3}+\frac{\lambda \phi_0^4}{3} )}{4\pi T}(r-r_h)+\cdots \,, \nn\\
A_z&=&A_{z0}+\frac{q^2A_{z0}\phi_0^2}{2\pi T}(r-r_h)+\cdots \,, \nn\\
\phi&=&\phi_0+\frac{\phi_0 \big(q^2A_{z0}^2+h_0(m^2+\lambda \phi_0^2) \big)}{4h_0 \pi T}(r-r_h)+\cdots \,. \nn
\eea
The independent parameters are $T, r_h, f_0, h_0, A_{z0},\phi_0$, and $\beta$. Using the above scaling symmetries, we can reduce these seven free parameters to $T, A_{z0}, \phi_0$, and $\beta$, which correspond to three dimensionless parameters ($\frac{M}{b}, \frac{T}{b}, \frac{\beta}{b}$) in the dual field theory. For given ($\frac{M}{b}, \frac{T}{b}, \frac{\beta}{b}$), the numerical solutions of the above equations of motion can be obtained by the shooting method.

%%%%%%%%%%%%%%%%%%%%%%%%%%%%%

\end{document}